\documentclass[11pt,preprint]{aastex}

\usepackage{graphicx}        
\usepackage{txfonts}
\usepackage{color}
\usepackage{float}

\newcommand{\etal}{{\it et al.}}

\newcommand{\be}{\begin{equation}}
\newcommand{\ee}{\end{equation}}
\newcommand{\bea}{\begin{eqnarray}}
\newcommand{\eea}{\end{eqnarray}}
\newcommand{\ba}{\begin{array}}
\newcommand{\ea}{\end{array}}

\newcommand{\bit}{\begin{itemize}}
\newcommand{\eit}{\end{itemize}}
\newcommand{\ben}{\begin{enumerate}}
\newcommand{\een}{\end{enumerate}}

\begin{document}

\title{A Solar Dynamo Model Driven by Mean-Field Alpha and Babcock-Leighton Sources: Fluctuations, Grand-Minima-Maxima and Hemispheric Asymmetry in Sunspot Cycles}

\author{D. Passos\altaffilmark{1, 2, 3, A}, D. Nandy\altaffilmark{4, 5, B}, S. Hazra\altaffilmark{4}, I. Lopes\altaffilmark{1, 2}}

\altaffiltext{1}{CENTRA, Departmento de  F\'\i sica, Instituto Superior T\'ecnico, Av. Rovisco Pais, 1049-001 Lisboa, Portugal}
\altaffiltext{A}{dariopassos@ist.utl.pt}

\altaffiltext{2}{Departamento de F\'\i sica, Universidade de \'Evora, Col\'egio Ant\'onio Luis Verney, 7002-554 \'Evora, Portugal}

\altaffiltext{3}{GRPS, D\'epartment de Physique, Universit\'e de Montr\'eal, C.P. 6128, Centre-ville, Montr\'eal, Qc, Canada  H3C-3J7}

\altaffiltext{4}{Department of Physical Science, Indian Institute of Science Education and Research Kolkata,  Mohanpur 741252, West Bengal, India}

\altaffiltext{5}{Center of Excellence in Space Sciences India, IISER Kolkata, Mohanpur 741252, West Bengal, India}
\altaffiltext{B}{dnandi@iiserkol.ac.in}


\begin{abstract}
Extreme solar activity fluctuations and the occurrence of solar grand minima and maxima episodes, such as the Maunder minimum and Medieval maximum are well established, observed features of the solar cycle. Nevertheless, such extreme activity fluctuations and the dynamics of the solar cycle during Maunder minima-like episodes remain ill-understood.
We explore the origin of such extreme solar activity fluctuations and the role of dual poloidal field sources, namely the Babcock-Leighton mechanism and the mean-field $\alpha$-effect in the dynamics of the solar cycle. We mainly concentrate on entry and recovery from grand minima episodes such as the Maunder minimum and the  dynamics of the solar cycle, including the structure of solar butterfly diagrams during grand minima episodes.
We use a kinematic solar dynamo model with a novel set-up in which stochastic perturbations force two distinct poloidal field $\alpha$ effects. We explore different regimes of operation of these poloidal sources with distinct operating thresholds, to identify the importance of each. The perturbations are implemented independently in both hemispheres which allows one to study the level of hemispheric coupling and hemispheric asymmetry in the emergence of sunspots.
From the simulations performed we identify a few different ways in which the dynamo can enter a grand minima episode. While fluctuations in any of the $\alpha$ effects can trigger intermittency, in keeping with results from a mathematical time-delay model, we find that the mean-field $\alpha$-effect is crucial for the recovery of the solar cycle from a grand minima episode which a Babcock-Leighton source alone, fails to achieve. Our simulations also demonstrate a range of hemispheric asymmetry, including ``failed grand minima'' in one hemisphere while the other remains quiescent, to, both hemispheres exhibiting grand minima like conditions.
We conclude that stochastic fluctuations in two interacting poloidal field sources working with distinct operating thresholds is a viable candidate for triggering episodes of extreme solar activity and that the mean-field $\alpha$-effect capable of working on weak, sub-equipartition fields is critical to the recovery of the solar cycle following an extended solar minimum. Based on our results, we also postulate that solar activity can exhibit significant parity shifts and hemispheric asymmetry, including phases when one hemisphere is completely quiescent while the other remains active, to, successful grand minima like conditions in both hemispheres.
\end{abstract}


\maketitle


\section{Introduction}
For hundreds of years, mankind has kept a close eye on the evolution of our star. Since the advent of the telescope, solar activity as been monitored systematically, creating one of the longest continuous observation programs in the history of science \citep{Owens2013}, the sunspot number measurements. These solar features still continue to be used as the most common proxy for solar magnetic activity as its appearance in the Sun's surface is thought to be connected with the build up of large scale, coherent magnetic fields in the solar convection zone (SCZ) (mainly in the azimuthal direction, the so called toroidal magnetic field component). This field component is produced by a magnetohydrodynamic (MHD) dynamo that converts a portion of the solar differential rotation's kinetic energy into magnetic energy \citep{Parker1955}. This process takes place in a thin shear layer at the bottom of the SCZ, the \textit{tachocline}, between the turbulent convection zone and the underlying
stable overshoot layers. The most common theoretical framework used to explain the origin and evolution of the large
scale solar magnetic field is \textit{Mean-Field Dynamo Theory} \citep{Steenbeck1966}. Models based in this theory,
solve a simplified version of the MHD induction equation, usually in spherical-polar coordinates under some physically inspired simplifications such as axisymmetry and spatio-temporal averages of the magnetic field (see review by \citet{Charbonneau2010}). Considered by some, as the most successful of these models, \textit{Babcock-Leighton (BL) flux transport dynamo models} \citep{Wang1991, Durney1995, Dikpati1999, Nandy2002} have become very useful tools in explaining the main features of the solar cycle such as its periodicity, parity, the equatorward migration pattern of sunspots and the phase difference between the toroidal and dipolar magnetic fields. These models incorporate the solar meridional circulation and emulate the effect of the
decay of active regions for the production of poloidal field based on the original ideas of Babcock and Leighton \citep{Babcock1955, Leighton1969}.

For many decades, the importance of the BL surface mechanism was overshadowed by another type of poloidal field source, the mean-field $\alpha$-effect (or Parker's $\alpha$ effect) in mean-field $\alpha\Omega$ dynamos. In these type of dynamo models, the source for transforming toroidal field into poloidal field is the action of helical turbulence in the twisting of flux tubes as they buoyantly rise through the SCZ \citep{Charbonneau2010}. The mean-field $\alpha$-effect is thought to be quenched by super-equipartition magnetic fields and effective only on relatively weaker fields. Works such as \citet{DSilva1993} raised questions regarding the effectiveness of this type of alpha effect by asserting that Joys' Law distribution (tilt angle dependence on latitude of bipolar sunspot pairs) requires very strong, super-equipartition toroidal field strengths; see also \cite{Fan2001}. Moreover, recent observational evidence points out that the BL mechanism is the predominant source for the solar cycle \citep{DasiEspuig2010}. Following these developments the stakes seem to be loaded against the mean-field $\alpha$-effect mechanism. Nevertheless, recent, realistic 3D MHD simulations of the solar convection zone \citep{Kapyla2006, Brown2010, Racine2011}, show clear evidence that a mean field type $\alpha$-effect should exist in the bulk of the convection zone. The question is whether this mean-field $\alpha$ plays a dynamically important role in the solar cycle?

Important constraints on dynamo models have emerged in the last decade
based on long term reconstructions of solar activity using cosmogenic isotopes \citep{Miyahara2006, Usoskin2007, Usoskin2008}. These reconstructions show that besides the typical variability observed in the 11 year solar cycle, there are extended periods of time in which the Sun behaves in a very unusual way. Solar magnetic activity seems to exhibit \textit{intermittency} (fluctuations and episodes of stronger than normal and quieter than normal activity) -- as it is usually defined in the jargon of Dynamic Systems. This means that there are periods where the normal cyclic activity ceases to exist and the Sun enters a \textit{grand minimum} period. During these grand minima, the Sun's exhibits a very low number (or lack) of sunspots in its photosphere and it is believed that the overall magnetic field also becomes very weak. Thanks to recent data, e.g. \citet{Steinhilber2012}, we now know that these periods are much more common than previously thought and therefore they should be considered as a constraint on any model of the solar cycle. The most famous of these quiescent periods of activity is called the Maunder minimum and took place between 1645 and 1715 \citep{Maunder1904, Eddy1976}. This period of low solar activity coincided with a period of very low temperatures in the Earth's northern hemisphere during the last part of the Little Ice Age. This and other interesting correlations between solar activity and Earth's climate have been found in the past \citep{Haigh2007}. A vitally important aspect of the grand minima enigma is to understand how the solar dynamo reawakens from these quiescent phases and the dynamics leading to, during, and exit from grand minima phases.

\smallskip

Over the years some authors have focussed their efforts in modelling and understanding the mechanisms behind the observed short and long term variability (e.g. amplitude variations from cycle to cycle and grand minima-maxima) of the solar cycle.
Two major approaches are usually used in these studies, one that looks for dynamical effects and a second that studies the impact of stochasticity in the system. The dynamical mechanism is based non-linear interactions between the magnetic field and plasma flows. One of the most important ingredients in any dynamo model is the differential rotation. So, it is only natural to assume that perturbations in this strong, large scale flow can induce fluctuations in activity. The so called Malkus-Proctor effect, i.e., the feedback of the large scale magnetic field on the rotation is such an example, wherein, strong magnetic fields quench the very flow which sustains dynamo action. This results in a weakening of magnetic activity, subsequently to which the Reynolds stresses restore the differential rotation to its initial state, thereby resuming activity \citep{Tobias1997, Brook2002, Bushby2006}. While this remains a plausible scenario for activity modulation, recent analysis of global 3D MHD simulations of solar convection that present dynamo action, indicates that modulation effects on the differential rotation could be much smaller than previously thought \citep{Beaudoin2012}. Other examples of dynamical processes are the modulation of the meridional circulation by the magnetic field as seen in dynamo models working in the non-kinematic regime \citep{Rempel2006, Passos2012} and the time delay dynamics that are inherent in the dynamo system with spatially segregated source-layers \citep{Charbonneau2005, Wilmot-Smith2006, Jouve2010}.

The modelling approach based on stochastic fluctuations is motivated by the fact that the solar dynamo resides in a very turbulent environment, i.e., the SCZ. So it is highly plausible that some of the intervening physical mechanisms are affected by "noise" imparted by random processes that are the basis of turbulent convection. This is usually modeled as stochastic fluctuations around an average typical value of some model's ingredient such as the $\alpha$ effect \citep{Hoyng1988, Choudhuri1992, Hoyng1993, Schmitt1996, Hoyng1994, Charbonneau2004, Usoskin2008} or the meridional circulation \citep{Charbonneau2000, LopesPassos2009, Karak2010}. Even if noise is introduced in the system at correlation time scales which are much shorter than the solar cycle timescale they tend to induce modulations on time scales ranging from decades to centennia. These models are also very robust because they can handle a wide range of noise levels. Some of these "stochastically forced models" have been very  successful in providing a simple explanation for the observed short term solar variability, and a few of them also display grand minima-like episodes as well. For the interested reader, a comprehensive review on modulation mechanisms of the solar dynamo can be found in \cite{Tobias2002}.

Herein, we base our study on a stochastically forced, kinematic, BL dynamo model. Inspired by the study of \citet{Hazra2014} (henceforth \textsf{Paper I}), here we explore the role of stochastically forced, dual poloidal field sources in the context of grand minima and maxima in solar activity. In \textsf{Paper I} the authors present a novel solution on how the dynamo re-emerges from a grand minima episode. In that paper they use a time-delay low order dynamo model to explore the impact of having two $\alpha$-effects working simultaneously, one effective on weak toroidal fields (mean field $\alpha$) and another effective on stronger fields (BL $\alpha$). They demonstrate that the weak mean field effect has a crucial role in driving the dynamo out of a grand minimum and postulate the usage of both an upper and lower thresholds for the BL $\alpha$ in order to avoid ``fake'', physically incorrect, recovery from grand minima episodes. Here we follow this postulate, test the original idea and also study additional dynamics associated with extreme solar activity with a spatially extended solar dynamo model with solar-like differential rotation and other physically inspired parameterizations. We introduce a secondary, mean field $\alpha$-effect in the bulk of the convection zone in our BL flux transport model and explore the consequent dynamics in the stochastically forced regime. Our results support the main conclusions presented in \textsf{Paper I} and we explore the subject further. In the following section we discuss the base model and show how fluctuations in the surface BL $\alpha$ effect induce the dynamo to enter a grand minima-like state. In the subsequent section we add the secondary mean field $\alpha$-effect to the model, discuss the modifications introduced to the parametrization of both $\alpha$-effects and demonstrate scenarios wherein stochastic fluctuations trigger grand maxima and minima behaviours and self-consistent recovery from them. We also analyze the resultant butterfly diagrams and demonstrate that hemispheric coupling during quiescent phases can range from strong asymmetry to near simultaneous occurrence of quiescent phases in both the Northern and Southern solar hemispheres. We end with a concluding section discussing the implications of our results and putting them in the broader context.


\section{The model}

The axisymmetric dynamo equations (\ref{eq:1}) and (\ref{eq:2}) are solved in the kinematic regime using a modified version of the \texttt{Surya - code} \citep{Nandy2002, Chatterjee2004}. This code solves the equations
\bea
    \frac{\partial A}{\partial t} + \frac{1}{s}(\mathbf{v}\cdot\nabla)(s A) &=& \eta_p
    \left( \nabla^2 - \frac{1}{s^2}\right)A + \alpha B\,\,\,, \label{eq:1}\\
    \frac{\partial B}{\partial t} + \frac{1}{r}\left[\frac{\partial}{\partial r}(r v_r B)
    + \frac{\partial}{\partial \theta}(v_\theta B)\right] &=& \eta_t
    \left( \nabla^2 - \frac{1}{s^2}\right)B \nonumber \\
    && +\, s ((\nabla \times [A(r,\theta)\mathbf{e}_\phi])\cdot\nabla)\Omega \nonumber \\
    && +\, \frac{1}{r}\frac{d\eta_t}{dr}\frac{\partial}{\partial r}(rB)\,\,\,,
    \label{eq:2}
\eea
where $s=r \sin(\theta)$, $\mathbf{v}$ is the meridional flow, $\Omega$ the differential rotation and $\alpha$ is a surface source term that emulates the BL mechanism. This numerical model assumes different magnetic diffusivities for the poloidal and toroidal field componenets, $\eta_p$ and $\eta_t$ respectively.
The use of two distinct diffusivity profiles in the Surya code is inspired by the reasoning that strong magnetic fields will suppress turbulence.  Hence it is expected that in the strong field regions where the toroidal field resides -- turbulent diffusivity will be much less effective, while in the weak field regions occupied by the poloidal component -- the turbulent diffusivity will be higher (closer to values predicted by mixing length theory). This is a
relatively simple, although effective technique for numerically capturing this physics.
 The model also has a built-in buoyancy algorithm that searches for toroidal field exceeding a certain threshold ($10^5$G by default) at the base of the SCZ (at $r=0.71R_\odot$) at certain time intervals, removes half of it and deposits this at near-surface layers where the BL poloidal source is located. This is done in order to emulate the eruption of magnetic flux tubes. For most of our simulations, equations (\ref{eq:1}) and (\ref{eq:2}) are solved in a 256$\times$256 grid between $R_b=0.55R_\odot$ $< r <$ $R_\odot$ and 0 $< \theta <$ $\pi$ under appropriate
boundary conditions.

\begin{figure*}[htb]
    \centering
    \includegraphics[width=6.5 cm]{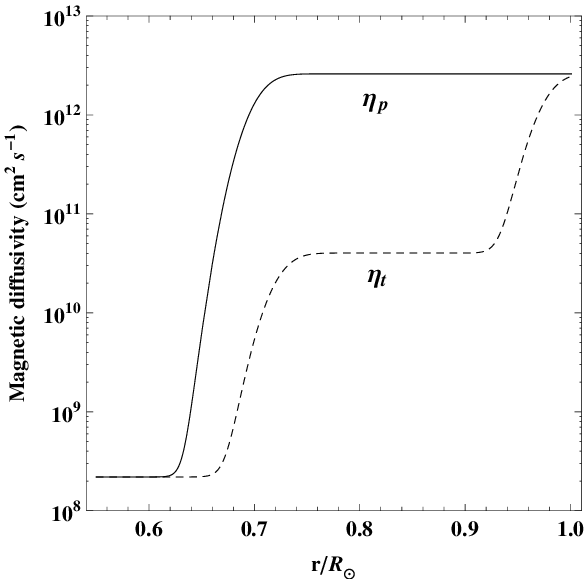}\,\,\,\,\,\,
    \includegraphics[width=6.3 cm]{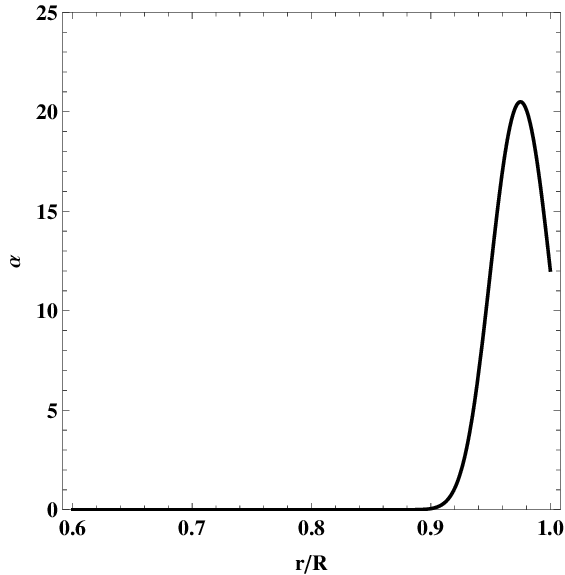}
    \caption[Surya's magnetic diffusivities and $\alpha$]{Left panel shows the toroidal (dashed)
    and poloidal (solid) magnetic diffusivities profiles. In the right the radial profile of
    $\alpha$ coefficient that is used to model the near-surface BL poloidal source (in m s$^{-1}$).}
    \label{fig:1}
\end{figure*}

This code also uses a differential rotation profile, $ \Omega(r,\theta)$, that is an analytical fit to the solar internal rotation data provided by Helioseismology. The meridional circulation, $\mathbf{v}$, has a one cell per hemisphere configuration, with latitudinal flows drifting toward the poles in the surface layers and towards the equator just below the tachocline. The amplitude of the surface component of this flow, at mid latitudes, is defined by $v_0=-29$ m s$^{-1}$ (in the standard version of the code). Detailed description of mathematical parametrization of these quantities including their justifications, for this well tested dynamo code, are available elsewhere \citep{Nandy2002, Chatterjee2004, Yeates2008}.
The standard version of this code implements a surface \emph{$\alpha$ effect} that mimics the BL mechanism and is parameterized through
\be
    \alpha(r,\theta) = \alpha_0\frac{ \cos \theta }{4} \left[1+\textrm{erf}
    \left( \frac{r-r_1}{d_1}\right)\right] \times\ \left[1-\textrm{erf}
    \left( \frac{r-r_2}{d_2}\right)\right]\,\,\,\,,
    \label{c6.eq-alphastandard}
\ee
where $r_1=0.95R_\odot$, $r_2=R_\odot$, $d_1=d_2=0.025R_\odot$ are scaling factors and $\alpha_0=25$ m s$^{-1}$ is the amplitude of the $\alpha$-effect.

\begin{figure*}[htb!]
        \centering
        \includegraphics[width=16 cm ]{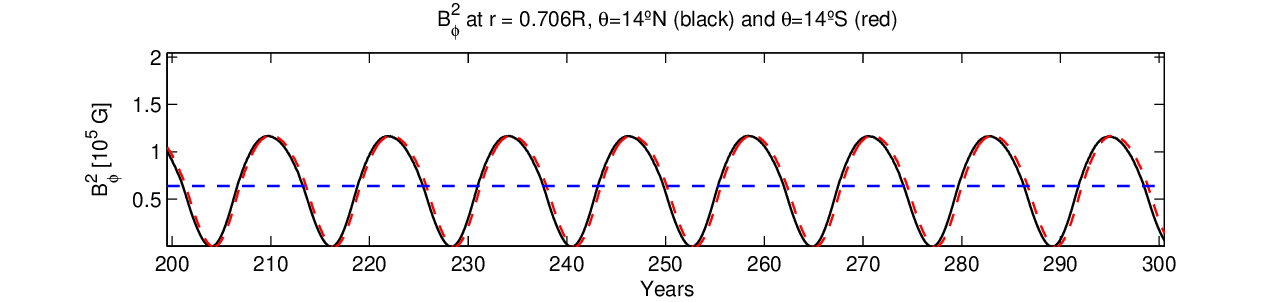}
        \includegraphics[width=16 cm]{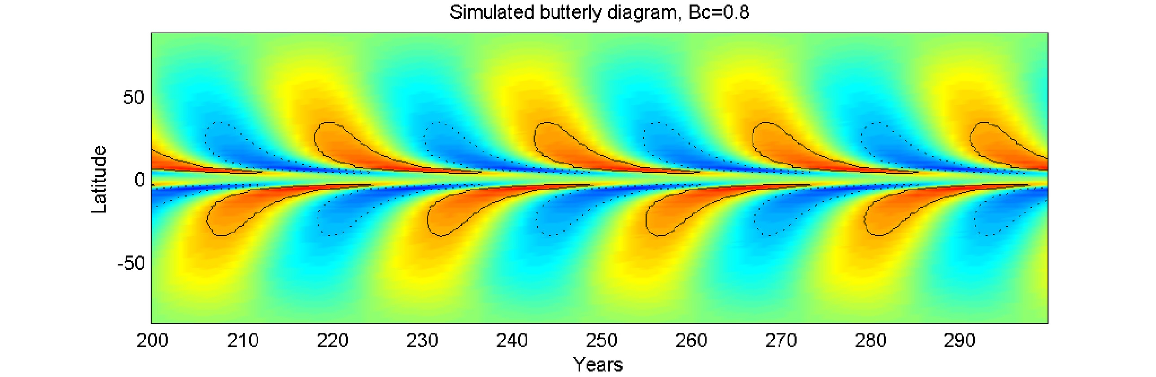}
        \caption{Top panel represents the toroidal field amplitude,
        $B_\phi^2$ at $r=0.706 R_\odot$ at $\theta=14^\circ$ North (solid, black) and
        South (dashed, red).
        The dashed line represents the buoyancy threshold. The corresponding analog to a
        butterfly diagram is presented below, for the field at the same depth.
        Blue represents negative field, red positive fields and green an average of zero.
        The contours enclose the areas where eruptions occur (B$_\phi>$B$_c$).}
        \label{fig:2}
\end{figure*}

In this study we use as a proxy for the solar cycle the amplitude of the toroidal field component given by $B_\phi^2$ just above tachocline depth and at usual active latitudes ($\sim$ 14$^\circ$). Figure \ref{fig:2} shows a reference solution depicting the evolution of the toroidal field in both hemispheres (red and black) just above the tachocline depth, at $r=0.706 R_\odot$ and the corresponding synoptic representation (akin to a butterfly diagram) where the positive (negative) erupted field is identified by solid (dashed) contour lines. This reference solution was obtained using $\alpha_0$=27 m s$^{-1}$, and the threshold for the buoyancy algorithm, B$_c=8\times10^4$ G. All other parameters are set to the default values as in the public version of the Surya dynamo code.

\section{Modelling grand minima episodes with stochastic fluctuations in the $\alpha$-effect}

A physical mechanism that has been proposed to be responsible for grand minima is the stochastic nature of the poloidal source -- traditionally the $\alpha$ effect (\citet{Choudhuri1992}, \citet{Charbonneau2000}, \citet{Proctor2007}, \citet{Brandenburg2008a}, \citet{Moss2008}). As noted before, as flux tubes rise through the CZ they are acted upon by the Coriolis force that imparts them the tilt angle observed in bipolar sunspot pairs. The values for the average tilt angle presents a uniform distribution centered around 6$^\circ$ \citep{Howard1991, DasiEspuig2010}. The idea behind this tilt dispersion is that during the buoyant rise, the turbulent buffeting of the tubes adds a random component to the tilt angle, contributing to the scatter in the distribution. In turn, this scattering in the tilt angles of active regions will contribute to variations in the efficiency of the surface BL mechanism. This phenomena (variations in the BL poloidal source amplitude) can, in principle, be modeled by introducing stochastic fluctuations in the amplitude of the BL $\alpha$-effect.

With this in mind we now redefine this coefficient by splitting it into a constant and a fluctuating part so that, $\alpha=\alpha_0 + \alpha'\sigma(t,\tau)$, where $\sigma$ is a function (with random values between -1 and 1) that depends on the time, $t$, on the correlation time for fluctuations, $\tau$, and $\alpha'$ the fluctuations' amplitude. Flux tube simulations suggest that their rise time through the CZ is of the order of a few months (e.g. \citet{Caligari1995}) and we know that surface flows takes on the order of months to redistribute the sunspot flux. Therefore, we choose $\tau=6$ months. As a first test, we implement fluctuations at 100\% level in $\alpha$ independently in both hemispheres. Since the fluctuations' levels can be difficult to estimate, the amplitudes we use are motivated by previous works (\cite{Charbonneau2000, Brandenburg2008b}), and on the eddy velocity distributions present in some highly turbulent global 3D MHD simulations of solar convection (e.g. \cite{Racine2011, Passos2012}). Also, significant fluctuation in the Babcock-Leighton source term is expected because of the large variation in the tilt angle distribution of bipolar solar active regions \citep{DasiEspuig2010}.

An important result from this experiment is the fact that fluctuations on a time scale much smaller (few months) than the solar cycle itself (approx. 11 years) can induce variability on the time scale of decades. Although the main cyclic activity persists, this shows how susceptible the system is to forcing factors.

\begin{figure*}[htb!]
        \centering
        \includegraphics[width=16 cm]{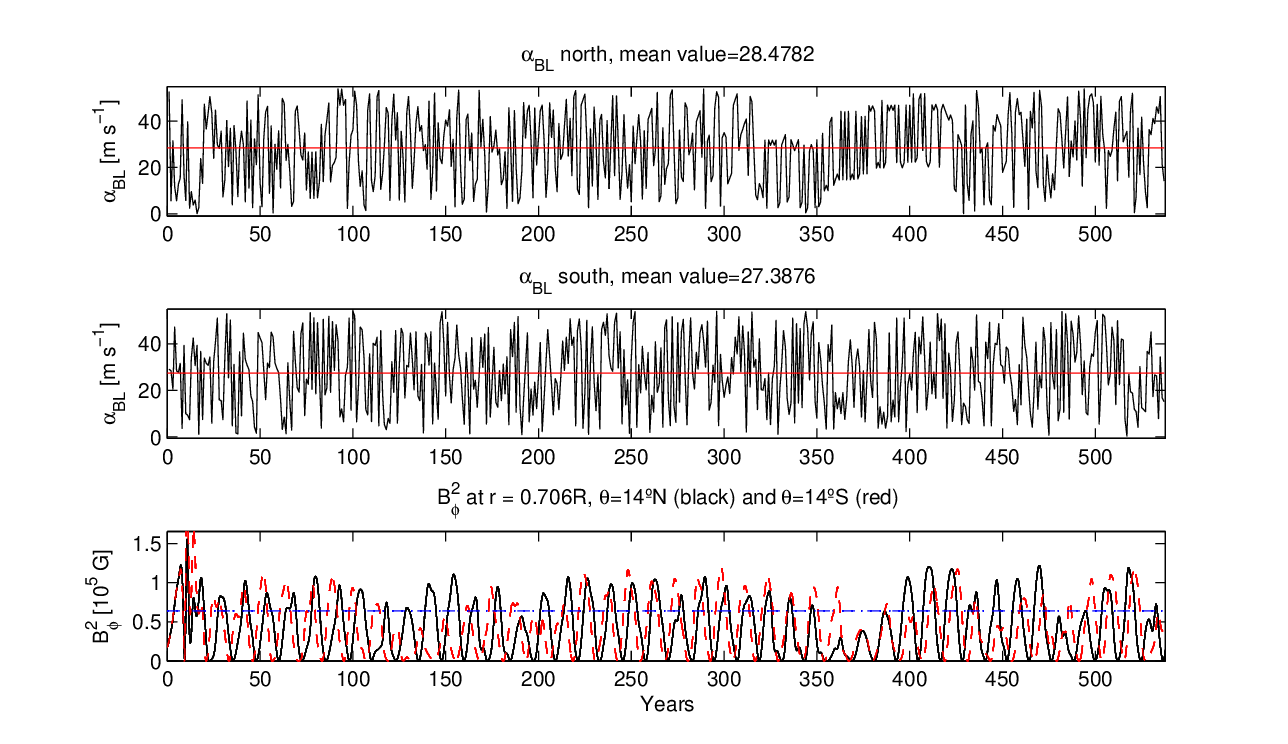}
        \includegraphics[width=16 cm]{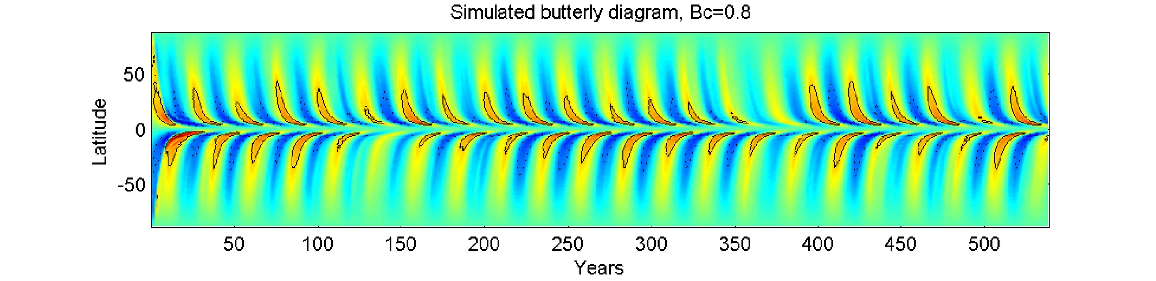}
        \caption{The two top panels show the fluctuating amplitude of $\alpha$ in the two hemispheres. The toroidal field amplitude at $r=0.706 R_\odot$ and at $14^\circ$ N (black) and $14^\circ$ S (red) is represented in the middle panel.
        The corresponding analog to a butterfly diagram is presented in the bottom panel.}
        \label{fig:3}
\end{figure*}

As an example of the obtained behaviour we present the result of one of the simulations in Figure \ref{fig:3}. In this figure we can identify periods (e.g. between 110 and 140 or between 340 to 390) where two or three cycles are suppressed in the synoptic representation. These periods would correspond to grand minima, in which the dynamo is still operating but in a regime where the produced fields are not strong enough to produce surface eruptions, i.e., the field amplitude stays below the buoyancy threshold (blue dashed line in the graphic).

\subsection{Two poloidal sources at play}

Although the previous simulation setup with fluctuations in a dynamo system with a Babcock-Leighton poloidal source alone, represents a scenario for entry and exit from grand minima like conditions, as has been demonstrated in earlier simulations \citep{Charbonneau2000, Karak2010, Choudhuri2012}, the physics of the recovery from a grand minimum in such model setups is questionable \citep{Hazra2014}. The problem lies with the parametrization of the Babcock-Leighton mechanism itself. Since the main idea behind this mechanism\footnote{Henceforth we will use the subscript \emph{BL} to denote the Babcock-Leighton $\alpha$-effect and \emph{MF} for the classical mean field $\alpha$-effect.} is the decay of active regions, one should ensure that this $\alpha_{BL}$ is only acting on strong fields that have buoyantly erupted and not on weaker fields that might reach the surface layers by advection or diffusion. This implies that a lower threshold in the $\alpha_{BL}$ must be introduced to prevent this (\citet{Nandy2002b, Charbonneau2005a, Hazra2014}). Therefore we introduce a new parametrization in the $\alpha_{BL}$ that depends of the field intensity in the following way

\bea
    \alpha_{BL} &=& \alpha_{0BL}\frac{ \cos \theta }{4} \left[1+\textrm{erf}
    \left( \frac{r-r_1}{d_1}\right)\right] \times\ \left[1-\textrm{erf}
    \left( \frac{r-r_2}{d_2}\right)\right] \nonumber \\
    && \times \, a_1\, \left[ 1 + \textrm{erf}
    \left(\frac{B_\phi^2 - B_{1lo}^2}{d_3^2}\right)\right]\times
    \left[ 1 - \textrm{erf}\left(\frac{B_\phi^2 - B_{1up}^2}{d_4^2}\right)\right]\,\,\,,
    \label{c6.eq-alphaBL}
\eea
where $r_1$, $r_2$, $d_1$ and $d_2$ have the same values as defined for equation (\ref{c6.eq-alphastandard}), $\alpha_{0BL}$ controls the amplitude of the effect, $a_1=0.393$ is a normalization constant, $B_{1lo}=10^3$G corresponds to the lower threshold, $d_3=10^2$ G, $B_{1up}=10^5$ G is an upper threshold and $d_4=10^6$G. This formulation ensures that only the fields that reach the surface layers with magnitudes between $B_{1lo}$ and $B_{1up}$ contribute to the Babcock-Leighton mechanism.

In order to determine what would be an appropriate value for the lower threshold of $\alpha_{BL}$ in our model, $B_{1lo}$, we conducted the following experiment. We performed simulations where we let the model evolve until it reached a maximum concentration of toroidal field in the base of the convection zone (akin to a cycle maximum). In the following step we use this final state as the initial condition of a new simulation and, to ensure that there is no toroidal field in the main body of the convection zone, we added a condition to wipe out fields above $r=0.725 R_\odot$. In other words, we initiate a new simulation with a high concentration of toroidal field just below the tachocline, but none above. We switched off the buoyancy algorithm and let the dynamo solution evolve just in the presence of advective flows and diffusion. The time that the peak associated with this strong field ($\sim10^5$ G at $r=0.725 R_\odot$) takes to reach near surface layers at active latitudes, more specifically at a latitude $\theta=15^\circ$N and $r=0.97 R_\odot$, around the radius where $\alpha_{BL}$ peaks, is approximately 5 years. The field reaching this radius had an amplitude around 750 G. This implies that this field is being transported primarily by the meridional flow towards the surface. The diffusion time scale for the toroidal field is given by $\tau_t=\ell^2/\eta_t$, and direct substitution ($\ell \simeq 0.245 R_\odot$ and $\eta_t=4\times 10^{10}$ cm$^2$ s$^{-1})$ yields that $\tau_t$ is around 230 years. Even if we consider an average value of $\eta$ one order of magnitude higher, we still obtain a diffusion time several times higher than the advection time found.

This numerical experiment and associated physical insights point out that even if buoyant eruption is absent, such as that during solar grand minima episodes, in such model setups, toroidal field is still dredged up from beneath the base of the convection zone to the surface by meridional circulation. Even though this toroidal field takes 5 years to move up and is weak, it ends up at the near-surface region where the spatially distributed, $\alpha$-coefficient representing the Babcock-Leighton poloidal source is located. Not being able to distinguish whether this toroidal field represents a contribution from sunspot eruptions or due to dredging up of field due to meridional circulation, the BL $\alpha$ source ends up producing poloidal fields, in a manner which is un-physical and not in keeping with established ideas from the dynamics of magnetic flux tubes. It is well known, that if it takes many years for a toroidal flux tube to rise up, it will be completely shredded by turbulence and will not have the required tilt angle distribution to contribute to the BL mechanism. Thus, to circumvent this un-physical behaviour of such a model setup, a lower threshold of $B_{1lo}=10^3$ G is necessary in such kinematic dynamo models.

\begin{figure*}[htb]
        \centering
        \includegraphics[width=16 cm]{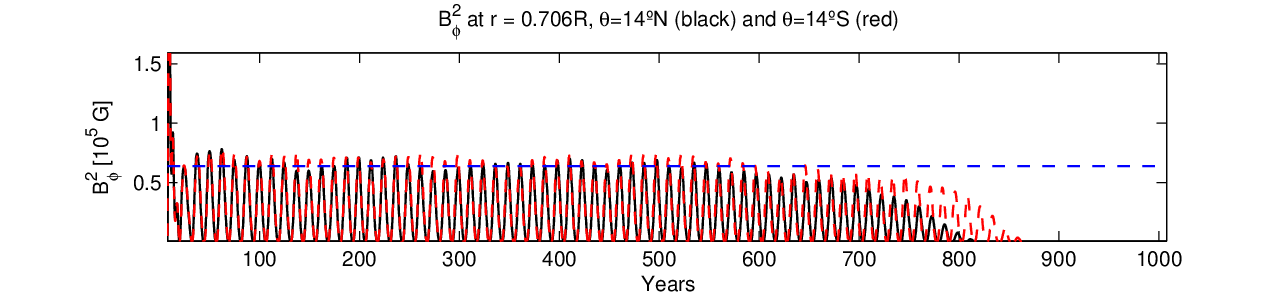}
        \caption{The toroidal field amplitude at $r=0.706 R_\odot$ at $\theta=14^\circ$ N (black) and S (red) is represented in the middle panel. In this model setup with both lower and upper thresholds and fluctuations in the BL mechanism, the dynamo never recovers once it enters a grand minima like phase.}
        \label{c6.fig-sim2}
\end{figure*}

The effect of this lower threshold when compared to the reference solution is to lower the amplitude of the solution's magnetic field and consequently decrease the number of eruptions and the latitudes at which they appear. The value of $\alpha_{BL}$=27 now corresponds to a value just above critical with eruptions first appearing at 25 degrees in latitude. This situation makes the dynamo solution specially susceptible to amplitude fluctuations in this coefficient. We show in Figure \ref{c6.fig-sim2} a dynamo solution which decays after half a century with just 10\% fluctuations in this redefined BL $\alpha$-coefficient with the lower threshold. The decay time of the solution scales inversely to the fluctuation levels. For the current setup, fluctuation levels between 50\% and 150\% induce the decay of the solution after a few decades, but for lower fluctuation level the the solution takes more time to decay. The observed decoupling between northern and southern hemispheres that appear in these solutions are also plausibly connected to the fluctuations levels.

Therefore, while a Babcock-Leighton mechanism with both upper and a lower operating thresholds allows for the possibility of entering a grand minima period, it does not seem to permit a self-consistent recovery from grand minima episodes; this conclusion is also supported by simulations with an independent, time-delay dynamo model presented in \textsf{Paper I}. Since the Sun has always emerged from such quiescent periods, one needs to come up with a solution that models this behaviour. The solution we present here is inspired by classical $\alpha \Omega$ models and its implementation is based on ideas discussed in \textsf{Paper I} in the context of a reduced dynamo model; i.e., we add a weak mean-field $\alpha$-effect working in conjunction with the surface $\alpha_{BL}$. This $\alpha_{MF}$ operates in the bulk of the convection zone and operates on weak flux tubes that are below a certain threshold and which do not contribute to the formation of sunspots (thus such an $\alpha$-effect could be operation even during grand minima phases).

\begin{figure*}[htb]
        \centering
        \includegraphics[height=5.0 cm]{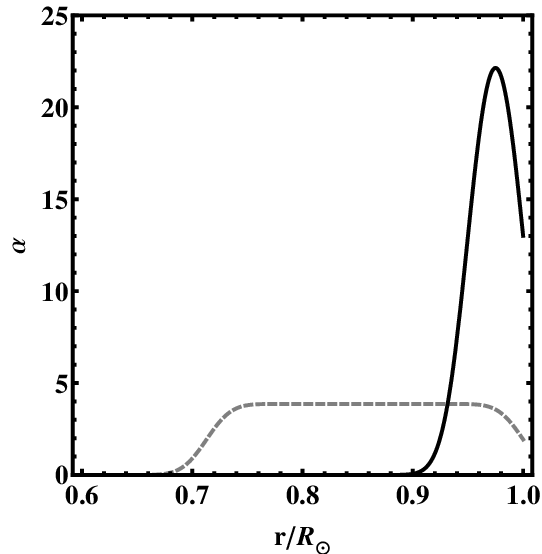}\,\,\,\,\,\,\,\,\,\,
        \includegraphics[height=5.15 cm]{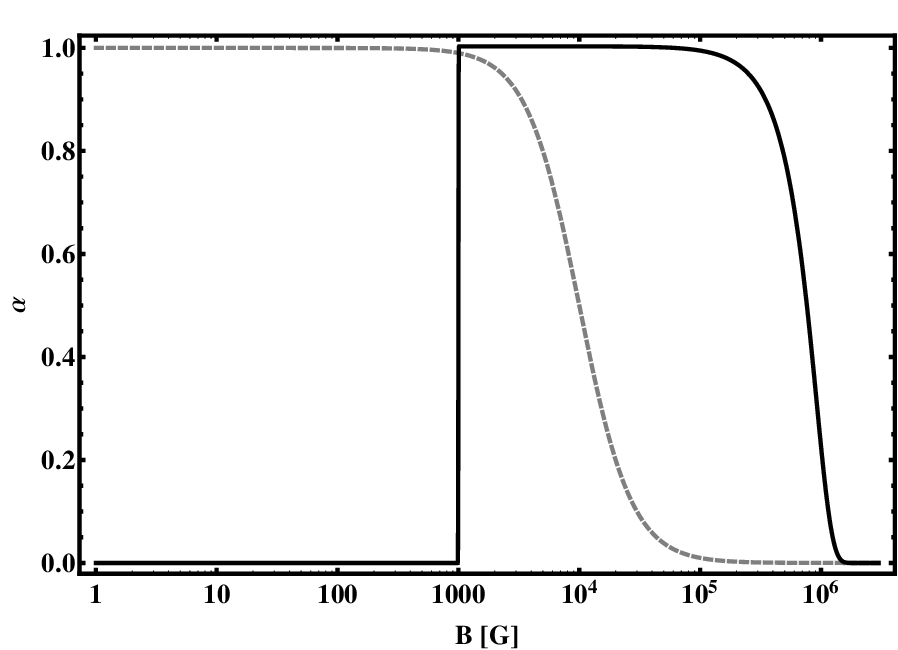}
        \caption{Radial profile (left panel) and quenching profile (right panel) for the Babcock-Leighton $\alpha_{BL}$ (black, solid curves) and for the mean field $\alpha_{MF}$ (gray, dashed curves). For better visual interpretation, the amplitude in the radial profile of the $\alpha_{MF}$ represented here is scaled up by one order of magnitude above the values used in the simulations.}
        \label{c6.fig-alphasprofiles}
\end{figure*}

We parameterize this mean-field $\alpha$ effect as

\bea
    \alpha_{MF} &=& \alpha_{0MF}\frac{ \cos \theta }{4} \left[1+\textrm{erf}
    \left( \frac{r-r_3}{d_1}\right)\right] \times\ \left[1-\textrm{erf}
    \left( \frac{r-r_4}{d_2}\right)\right] \nonumber \\
    && \times \, \frac{1}{1+\left(\frac{B_\phi}{B_{2up}}\right)^2}\,\,\,,
 \eea
where $\alpha_{0MF}$ is the amplitude of the effect, $r_3=0.713R_\odot$, $r_4=R_\odot$ and $B_{2up}=10^4$G is an upper threshold. The radial profile and quenching profiles for the two poloidal sources (parameterized through two $\alpha$ effects as described above) are presented in Figure \ref{c6.fig-alphasprofiles}. The combined final $\alpha$-effect acting in our system is then defined as $\alpha=\alpha_{BL}+\alpha_{MF}$, the first acting on strong fields that erupt buoyantly to the surface and the second acting on weaker fields that are either advected by meridional flow or diffuse out into the SCZ from the tachocline.

In the absence of any fluctuations, the solution obtained using this combined $\alpha$ with $\alpha_{BL}$=27, $B_{1lo}=10^3$ and $\alpha_{MF}=0.4$ is similar to the reference solution (Figure \ref{fig:2}) but with a slight decrease ($\sim$ 1 year) in the period of the cycle and eruptions starting from somewhat higher latitudes. They are qualitatively similar in other aspects.

\subsection{Stochastic fluctuations in the poloidal sources}

To test if the combination of these two poloidal sources can produce grand minima-like episodes (under the influence of stochastic fluctuations) and recover self-consistently from these episodes, we apply different levels of fluctuations to both $\alpha$ effects, separately and in conjunction to perform further simulations.

\begin{figure*}[htb]
        \centering
        \includegraphics[width=16 cm]{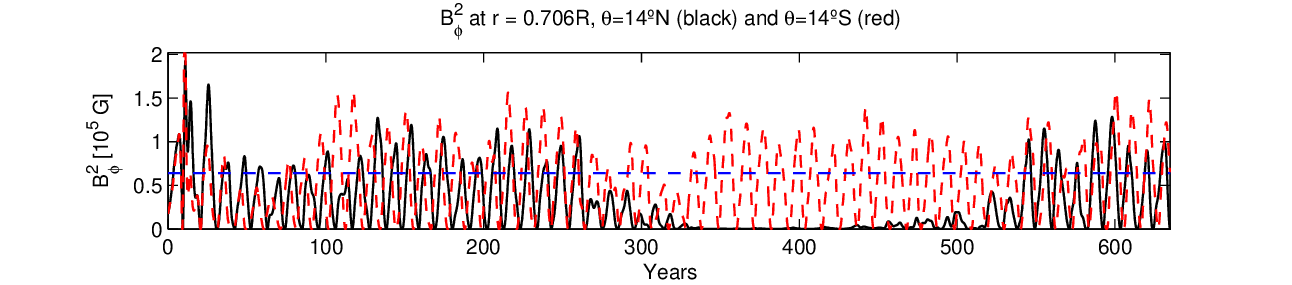}
        \includegraphics[width=16 cm]{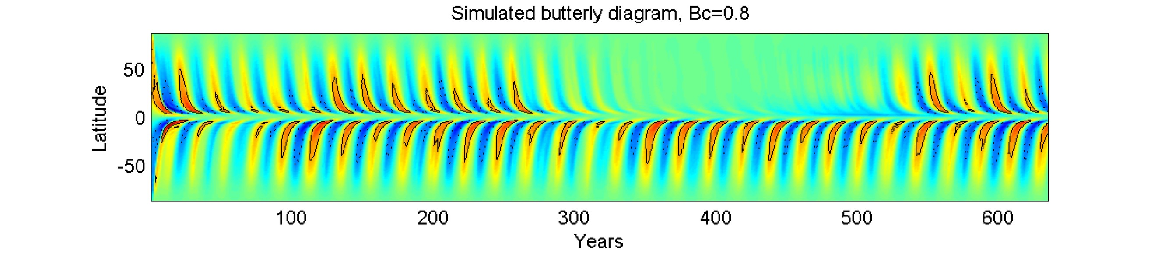}
        \caption{Simulation with $\alpha_{BL}=27$ + 100\% fluctuations, $\alpha_{MF}=0.4$. Between the year 280 and 540, there is a prolonged period without eruptions in the Northern hemisphere that is not accompanied by a similar behaviour in the South.}
        \label{c6.fig-sim12}
\end{figure*}

We start by repeating the simulation with 100\% fluctuations in the $\alpha_{BL}=27$ but now with the additional presence of $\alpha_{MF}=0.4$ (see Figure \ref{c6.fig-sim12}). We observe that there are periods where the field falls below the buoyancy threshold but after some time it regains strength and begins producing sunspot eruptions again. The underlying dynamics likely involves fluctuations in $\alpha_{BL}$ triggering grand minima periods, and the presence of $\alpha_{MF}$ facilitating recovery. Hints of such a dynamics between two interacting dynamo $\alpha$-effects were already present in \cite{Ossendrijver2000}, but where instead of the near-surface BL $\alpha$ effect, a buoyancy instability induced $\alpha$-effect in the overshoot layer was considered.

In these simulations we continue to observe the decoupling between the Northern and Southern hemispheres, especially in simulations with a high level of fluctuations in the Babcock-Leighton mechanism. Occasionally, small changes in the parity occur as well (deviations from dipolar parity). In some extreme cases, both hemispheres sustain dynamo action but in an almost independent manner. During the deepest phase of a grand minimum, the hemispheres loose synchronization but interestingly it is quickly regained after the system emerges from this quiescent period. There is a small, variable phase shift between the two hemispheres that was already present in Figure 3. This type of weak hemispheric decoupling during normal activity is plausibly due to the presence of fluctuations and is also observed in the solar butterfly diagram. Different levels of fluctuations in the $\alpha_{BL}$ between 25\% and 200\% were tested. We observe that the decoupling between hemispheres and the number of grand minima increases with the fluctuation level. Independent simulations in a different context also point out that hemispheric decoupling may be a phenomena that is naturally associated with, and symptomatic of grand minima episodes \citep{Olemskoy2013}. One also notes from Figure \ref{c6.fig-sim12} that based on the circumstances, one may have a quiescent minimum like phase in one hemisphere, but a failed minimum in the other hemisphere.

\bigskip

In the next batch of simulations we apply fluctuations just to the mean-field poloidal source $\alpha_{MF}$ using a correlation time of 1 year. The simulations with $\alpha_{BL}=27$ and $\alpha_{MF}=0.4$ + 100\% fluctuations returned very few (and short, a couple of cycles at the most) minima episodes. Increasing fluctuations to 200\% increases the number of minima but not their length. In order to test the importance of $\alpha_{MF}$ in the dynamics, we decrease the strength of the BL source to $\alpha_{BL}=21$ (barely supercritical). With this lower value, values of 100\% to 200\% fluctuations in the mean field produce longer minima episodes. Both hemispheres seem to evolve in a similar way although some variability between North and South is observed. Even for longer simulated periods, this coupled behaviour is maintained as well as the parity of the solution. This implies that the frequency of occurrence, and duration of grand minima episodes depend on the relative amplitudes of the two poloidal sources and the level of fluctuations in them. Moreover, it appears that the mean-field poloidal source -- distributed across the convection zone -- plays a more important role in the maintenance of hemispheric coupling and dipolar parity than the BL poloidal source.

\begin{figure*}[htb]
        \centering
        \includegraphics[width=15 cm]{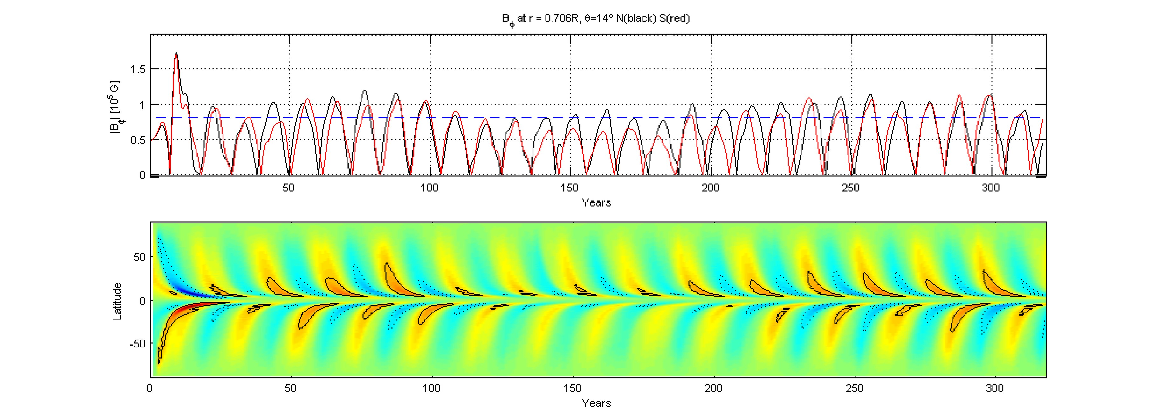}
        \caption{Simulation with $\alpha_{BL}=21$ and $\alpha_{MF}=0.4$ + 100\% fluctuations. Here, the two solar hemispheres appear to be well coupled.}
        \label{c6.fig-extra5-21}
\end{figure*}

\begin{figure*}[htb]
        \centering
        \includegraphics[width=15 cm]{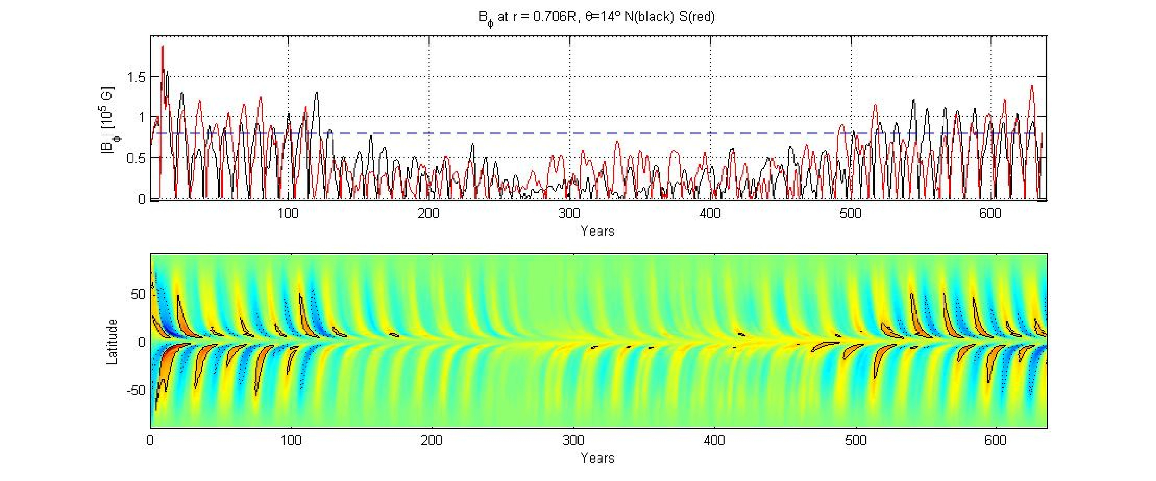}
        \includegraphics[width=12.6 cm]{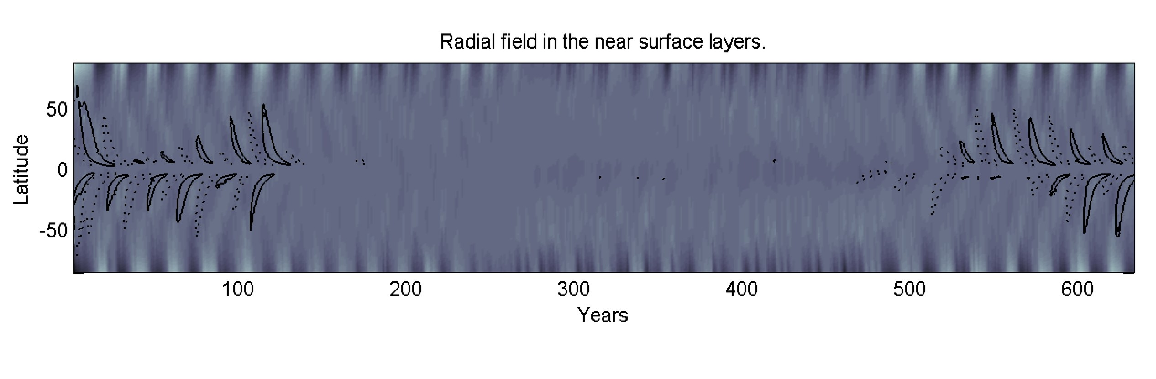}
        \caption{Simulation with $\alpha_{BL}=27$ + 100\% fluctuations and $\alpha_{MF}=0.4$ + 200\% fluctuations. Top panel depicts the toroidal field amplitude in North and South hemispheres, the middle panel is a butterfly diagram for the toroidal field at the base of the convection zone, and the bottom panel show the radial field near the surface (lighter and darker shades denotes positive and negative radial field, respectively). This solution, representative of simulations with intermittency in both mean-field and Babcock-Leighton poloidal sources, displays self-consistent entry and exit from grand minima episodes. The butterfly diagram is characterized by hemispheric asymmetry in activity, parity shifts and occasional sunspot eruptions during the minimum phase. The radial field evolution shows regular polarity reversals in the weak dipolar component of the magnetic field throughout the minimum in activity.}
        \label{c6.fig-sim15}
\end{figure*}

To complete the set of possible numerical experimentation, different levels of fluctuations can be applied to both the poloidal sources. The overall set of obtained solutions, not surprisingly, show a wide range of behaviours in terms of frequency and duration of grand minima, hemispheric decoupling and parity change. A representative solution from numerical simulations with intermittency in both poloidal sources is presented in Figure \ref{c6.fig-sim15}. Here we note a strong hemispheric asymmetry in the activity just prior to entering the grand minimum phase, continued hemispheric decoupling throughout the minimum phase and regaining of hemispheric coupling following recovery from the minimum. The period during the grand minimum is characterized by an extended period of low magnetic activity; however, there are occasional sunspot eruptions during this period and the solar polar field (dipolar component) continues its regular cycle of reversal albeit with weak field amplitudes. These features are in broad qualitative agreement with observed features of the solar Maunder minimum \citep{Beer1998, Miyahara2006}.

\section{Concluding Summary}

Solar dynamo models can in general explain the main features of the solar cycle. Explanations of solar cycle fluctuations, such as grand minima and maxima episodes should also be encompassed in such models. Radio-isotope records have shown that our Sun regularly goes through such quiescent periods and higher than normal activity phases. In order to study such activity modulation two different approaches are usually used. In the first approach, one utilizes the role of dynamic, non-linear feedback mechanisms in the solar dynamo \citep{Tobias1997, Brook2002, Bushby2006} including time delay dynamics \citep{Charbonneau2005, Wilmot-Smith2006, Jouve2010}. In the second approach, one utilizes stochastic forcing of the dynamo system motivated from the turbulent nature of the solar convection zone. In this work we follow the latter approach, in which using a stochastically forced, kinematic solar dynamo model based on the flux transport paradigm, we focus on solar cycle dynamics related to grand minima episodes.

 We may also point out that in this model, turbulent diffusion, and not meridional circulation, is the dominant process for transporting poloidal field in to the solar interior where the toroidal field is produced and stored. However, meridional circulation does play an important role in the near-surface poloidal field dynamics and latitudinal transport of toroidal fields deep in the interior. In this context it may be noted that another mechanism -- downward turbulent pumping of magnetic flux \citep{Tobias2001} -- is also expected to play an important role in solar cycle dynamics \citep{Guerrero2008, Nandy2012}. Moreover, turbulent pumping in a rotating medium may also contribute to the reprocessing of magnetic flux and thus be an effective mechanism for continuation of weak cycles; we expect that the current model set-up with a mean-field $\alpha$ effective on weak fields would encompass the consequences of such a mechanism implicitly, even in the absence of an explicit treatment of turbulent pumping.

Since the convection zone is highly turbulent, stochastic fluctuations arise naturally in the dynamo source terms. Motivated by the results obtained in a low-order dynamo system in \textsf{Paper I}, showing the importance of a dual source formalism in the context of self-consistent entry and exit from grand minima, we started by adding a lower quenching threshold to the parametrization of the BL poloidal source. Adding this low-threshold to the Babcock-Leighton mechanism ensures that this source only acts on strong fields that erupt to the surface and not on weak fields that ``leak'' into the convection zone by advection and diffusion. These weak fields do not produce sunspots, hence they do not participate in the Babcock-Leighton mechanism. We believe that this presents a more realistic approach to modeling the BL poloidal source. Although stochastic fluctuations (in $\alpha_{BL}$) in models without the lower threshold present grand minima like episodes, we argue that these are not real and are un-physical. When the threshold is included, the model returns solutions with a weaker amplitude of the magnetic field which are very susceptible to stochastic fluctuations and models with this physically justified BL source parametrization fail to recover from a grand minima as pointed out in \textsf{Paper I}.

An additional weak, classical mean-field effect, $\alpha_{MF}$, operating in the bulk of the convection zone is introduced in the model. This source acts on toroidal fields that are not strong enough to produce sunspots and are advected by the meridional flow (and diffuse) in to the convection zone. Our kinematic dynamo model with the dual poloidal sources presents typical dynamo solutions with emerging patterns similar to the classical sunspot butterfly diagram. This model presents more robust solutions when stochastic fluctuations are introduced to both $\alpha$-effects (separately or simultaneously). For several combinations in the parameter space, we observe behaviours analogous to grand minima, where the eruption of toroidal field at the surface is suppressed. However, with this dual poloidal source, physically consistent recovery from a grand minimum is possible as demonstrated in this spatially extended dynamo model. Thus, the results obtained herein corroborates and extends those already presented in \textsf{Paper I} (using a completely different modelling technique), which justifies the usage of appropriate low order models such as the time delay dynamo model used in \textsf{Paper I} to explore long-term phenomena which are computationally expensive for spatially extended dynamo models.

Two possible scenarios for entering and exiting a grand minimum were identified. The first is the case where small fluctuations in the surface $\alpha_{BL}$ are combined with fluctuations on $\alpha_{MF}$ in a model with a near critical solution. Since the field is near the limit necessary for the production of sunspots, fluctuations in $\alpha_{MF}$ can take it below the buoyancy threshold. These grand minima tend to be small with a duration of just a few cycles. The other case is in models with solutions above critical when both source terms are subjected to considerable levels of fluctuations (above 75\%). In this case, fluctuations in $\alpha_{BL}$ tend to make the toroidal field fall abruptly bellow the threshold necessary for the production of sunspot eruptions; this renders the Babcock Leighton source completely ineffective with consequent disruption of dynamo activity. The combination of different fluctuation levels also controls the onset of the grand minima (abrupt or gradual). In several cases we observe that one of the hemispheres emerges from a minimum before the other by a few cycles. This was also apparently the case during the exit from the Maunder Minimum \citep{Nagovitsyn2010} -- so this model behaviour is entirely plausible.

Indeed we find that fluctuations in $\alpha_{BL}$ tend to decouple the magnetic activity in the two hemispheres as observed in these spatially extended simulations. This is because in our model we allow for different randomly generated fluctuations in the Northern and Southern hemispheres -- which is expected in reality. Nevertheless for fluctuations of $\alpha_{BL}$ of the order of 25\% the asymmetry between hemispheric magnetic activity is small, but with increasing fluctuation levels, this asymmetry increases. We find that the mean field poloidal source plays a more important role in keeping the two hemispheres coupled, while fluctuations in the Babcock Leighton source tends to introduce parity shifts and decouple the two hemispheres. It is indeed possible, as demonstrated, that under certain conditions, one hemisphere may undergo minimum like conditions with no eruptions while there is a failed minimum in the other hemispheres where sunspots eruptions continue. Independent simulations point out that hemispheric asymmetry may be symptomatic of grand minima episodes \citep{Olemskoy2013} and \citet{Charbonneau2005} provides a possible physical explanation of the role of hemispheric coupling vis-a-vis solar grand minima episodes.

It is interesting to note that for a large number of combinations of poloidal source amplitudes and fluctuations, we see regular polar field reversals (in radial field evolution) even when the surface Babcock-Leighton mechanism remains ineffective over a long period during a grand minimum (due to the lack of bipolar sunspot eruptions). This behaviour in our model is due to the presence of the mean field poloidal source which maintains a low activity cycles. This result is supported by solar activity reconstructions based on cosmogenic isotopes which indicate regular polar field reversals occurred during the Maunder minimum \citep{Miyahara2004}.

One of the important aspects of extreme solar activity fluctuations is the observed statistics (frequency distribution) of solar grand minima and maxima episodes \citep{Usoskin2007}. These constraints are on the duration and waiting periods for minima and maxima episodes. While we have not addressed this aspect in our first comprehensive study with a dynamo driven by dual poloidal field sources, we point out that an independent study does focus on a comparative analysis of the observed statistics of grand minima with a kinematic dynamo model setup in a similar way \citep{Karak2013}. Another important aspect which we do not directly focus on here is the phase that immediately precedes a solar grand minimum and the defining  characteristics of this phase. Since such studies require detailed investigations, and significant computational investment in simulations, we defer this to the future.

\begin{acknowledgements}
We are grateful to the Ministry of Human Resource Development, Council for Scientific and Industrial Research, University Grants Commission and the Ramanujan Fellowship award of the Department of Science and Technology of the Government of India for supporting this research. D. Passos acknowledges support from the Funda\c{c}\~{a}o para a Ci\^{e}ncia e Tecnologia grant SFRH/BPD/68409/2010, CENTRA-IST, the University of the Algarve for providing office space and he is also grateful to IISER Kolkata for hosting him during the final phases of the research work presented here. We also acknowledge the referee for his critical comments which helped us improve an earlier version of our manuscript.
\end{acknowledgements}


\begin{thebibliography}{}

\bibitem[\protect\citeauthoryear{Babcock \& Babcock}{1955}]{Babcock1955}
Babcock, H. W., Babcock, H. D. 1955,
\apj, 121, 349

\bibitem[\protect\citeauthoryear{Beaudoin \etal} {2012}]{Beaudoin2012}
Beaudoin, P., Charbonneau, P., Racine, E., Smolarkiewicz, P. K. 2012,
\solphys, 282, 335

\bibitem[\protect\citeauthoryear{Beer \etal}{1998}]{Beer1998}
Beer, J., Tobias, S. M.,Weiss, N. O. 1998,
\solphys, 181, 237

\bibitem[\protect\citeauthoryear{Brandenburg \& Spiegel}{2008a}]{Brandenburg2008a}
Brandenburg, A., Spiegel, E. A. 2008a,
Astronomische Nachrichten, 329, 351

\bibitem[\protect\citeauthoryear{Brandenburg \etal}{2008b}]{Brandenburg2008b}
Brandenburg, A., Radler, K. -H. Rheinhardt, M., K\"{a}pyl\"{a}, P. J. 2008b,
\apj, 676, 740

\bibitem[\protect\citeauthoryear{Brooke \etal}{2002}]{Brook2002}
Brooke, J., Moss, D., Phillips, A. 2002,
\aap, 395, 1013

\bibitem[\protect\citeauthoryear{Brown \etal}{2010}]{Brown2010}
Brown, B. P., Browning, M. K., Brun, A. S., Miesch, M. S., Toomre, J. 2010,
\apj, 711, 424

\bibitem[\protect\citeauthoryear{Bushby}{2006}]{Bushby2006}
Bushby, P. J. 2006,
\mnras, 371, 772

\bibitem[\protect\citeauthoryear{Caligari \etal}{1995}]{Caligari1995}
Caligari, P., Moreno-Insertis, F., Sch\"{u}ssler, M. 1995,
\apj, 441, 886

\bibitem[\protect\citeauthoryear{Charbonneau \& Dikpati}{2000}]{Charbonneau2000}
Charbonneau, P., Dikpati, M. 2000,
\apj, 543, 1027

\bibitem[\protect\citeauthoryear{Charbonneau \etal }{2004}]{Charbonneau2004}
Charbonneau, P., Blais-Laurier, G., and St-Jean, C. 2004,
\apj, 616, 183

\bibitem[\protect\citeauthoryear{Charbonneau \etal}{2005}]{Charbonneau2005}
Charbonneau, P., St-Jean, C., Zacharias, P. 2005,
\apj, 619, 613

\bibitem[\protect\citeauthoryear{Charbonneau}{2010}]{Charbonneau2010}
Charbonneau, P. 2010,
Living Rev. Solar Phys., 7,  3

\bibitem[\protect\citeauthoryear{Chatterjee \etal}{2004}]{Chatterjee2004}
Chatterjee, P., Nandy, D., Choudhuri, A. R. 2004,
\aap, 427, 1019

\bibitem[\protect\citeauthoryear{Choudhuri}{1992}]{Choudhuri1992}
Choudhuri, A. R. 1992,
\aap, 253, 277

\bibitem[\protect\citeauthoryear{Choudhuri \& Karak}{2012}]{Choudhuri2012}
Choudhuri, A. R., Karak, B. B. 2012,
\prl, 109, 171103

\bibitem[\protect\citeauthoryear{Dasi-Espuig \etal}{2010}]{DasiEspuig2010}
Dasi-Espuig, M., Solanki, S. K., Krivova, N., Cameron, R., Pe\~{n}uela, T. 2010,
\aap, 518, 10

\bibitem[\protect\citeauthoryear{Dikpati \& Charbonneau}{1999}]{Dikpati1999}
Dikpati, M., Charbonneau, P. 1999,
\apj, 518, 508

\bibitem[\protect\citeauthoryear{Durney}{1995}]{Durney1995}
Durney, B. R. 1995,
\solphys, 160, 213

\bibitem[\protect\citeauthoryear{D'Silva \& Choudhuri}{1993}]{DSilva1993}
D'Silva, S., Choudhuri, A. 1993,
\aap, 272, 621

\bibitem[\protect\citeauthoryear{Eddy}{1976}]{Eddy1976}
Eddy, J. A. 1976,
Science, 192, 1189

\bibitem[\protect\citeauthoryear{Fan}{2001}]{Fan2001}
Fan, Y. 2001,
\apjl, 554, 111

\bibitem[\protect\citeauthoryear{Guerrero, de Gouveia Dal Pino}{2008}]{Guerrero2008}
Guerrero, G.,  de Gouveia Dal Pino, E. M. 2008,
\aap, 485, 267

\bibitem[\protect\citeauthoryear{Haigh}{2007}]{Haigh2007}
Haigh, J. D. 2007,
Living Rev. Solar Phys., 4, 2

\bibitem[\protect\citeauthoryear{Hazra \etal}{2014}]{Hazra2014}
Hazra, S., Passos, D., Nandy, D. 2014,
\apj, 789, 1

\bibitem[\protect\citeauthoryear{Hoyng}{1988}]{Hoyng1988}
Hoyng, P. 1988,
\apj, 332, 857

\bibitem[\protect\citeauthoryear{Hoyng}{1993}]{Hoyng1993}
Hoyng, P. 1993,
\aap, 272, 321

\bibitem[\protect\citeauthoryear{Hoyng}{1994}]{Hoyng1994}
Hoyng, P., Schmitt, D., Teuben, L. J. W. 1994,
\aap, 289, 265


\bibitem[\protect\citeauthoryear{Howard}{1991}]{Howard1991}
Howard, R.F. 1991,
\solphys, 136, 251

\bibitem[\protect\citeauthoryear{Jouve \etal}{2010}]{Jouve2010}
Jouve, L., Proctor, M. R. E., Lesur, G. 2010,
\aap, 519, 13

\bibitem[\protect\citeauthoryear{K\"{a}pyl\"{a} \etal}{2006}]{Kapyla2006}
K\"{a}pyl\"{a}, P. J., Korpi, M. J., Ossendrijver, M.,  Stix, M. 2006,
\aap, 455, 401

\bibitem[\protect\citeauthoryear{Karak}{2010}]{Karak2010}
Karak, B. 2010,
\apj, 724, 1021

\bibitem[\protect\citeauthoryear{Karak \& Choudhuri}{2013}]{Karak2013}
Karak, B., Choudhuri, A. R. 2013,
Res. Astron. Astrophys., 13, 1339


\bibitem[\protect\citeauthoryear{Leighton}{1969}]{Leighton1969}
Leighton, R. B. 1969,
\apj, 156, 1

\bibitem[\protect\citeauthoryear{Lopes \& Passos}{2009}]{LopesPassos2009}
Lopes, I., Passos, D.  2009,
\solphys, 257, 1

\bibitem[\protect\citeauthoryear{Maunder}{1904}]{Maunder1904}
Maunder, E. W. 1904,
\mnras, 64, 747

\bibitem[\protect\citeauthoryear{Miyahara, H. \etal}{2004}]{Miyahara2004}
Miyahara, H., Masuda, K., Muraki, Y., Muraki,  Y., Furuzawa, H., Menjo, H., Nakamura, T. 2004,
\solphys, 224, 317

\bibitem[\protect\citeauthoryear{Miyahara, H. \etal}{2006}]{Miyahara2006}
Miyahara, H., Masuda, K., Muraki, Y., Kitagawa, H., Nakamura, T., 2006,
\jgr, 111, A03103

\bibitem[\protect\citeauthoryear{Moss \etal}{2008}]{Moss2008}
Moss, D., Sokoloff, D., Usoskin, I., Tutubalin, V. 2008,
\solphys, 250, 221

\bibitem[\protect\citeauthoryear{Nandy \& Choudhuri}{2002}]{Nandy2002}
Nandy, D., Choudhuri, A. R. 2002,
Science, 296, 1671

\bibitem[\protect\citeauthoryear{Nandy}{2002b}]{Nandy2002b}
Nandy, D. 2002,
Ap \& SS, 282, 209

\bibitem[\protect\citeauthoryear{Nandy, Karak}{2012}]{Nandy2012}
Nandy, D., Karak, B. 2012,
American Astro. Soc., AAS Meeting 220, 521.18

\bibitem[\protect\citeauthoryear{Nagovitsyn \etal}{2010}]{Nagovitsyn2010}
Nagovitsyn, Y. A., Ivanov, V. G., Miletsky, E. V., Nagovitsyna, E. Y. 2010,
Astronomy Reports, 54, 476

\bibitem[\protect\citeauthoryear{Olemskoy \& Kitchatinov}{2013}]{Olemskoy2013}
Olemskoy, S. V., Kitchatinov, L. L. 2013,
\apj, 777, 71

\bibitem[\protect\citeauthoryear{Ossendrijver}{2000}]{Ossendrijver2000}
Ossendrijver, M. A. J. H. 2000,
\aap, 359, 364

\bibitem[\protect\citeauthoryear{Owens}{2013}]{Owens2013}
Owens, B. 2013,
Nature 495, 300

\bibitem[\protect\citeauthoryear{Parker}{1955}]{Parker1955}
Parker, E. N. 1955,
\apj, 121, 491

\bibitem[\protect\citeauthoryear{Passos \etal}{2012}]{Passos2012}
Passos, D., Charbonneau, P., Beaudoin, P. 2012,
\solphys, 279, 1

\bibitem[\protect\citeauthoryear{Proctor}{2007}]{Proctor2007}
Proctor, M. R. E. 2007,
\mnras, 382, 39

\bibitem[\protect\citeauthoryear{Racine \etal}{2011}]{Racine2011}
Racine, E., Charbonneau, P., Ghizaru, M., Bouchat, A., Smolarkiewicz, P. K. 2011,
\apj, 735, 46

\bibitem[\protect\citeauthoryear{Rempel}{2006}]{Rempel2006}
Rempel, M. 2006,
\apj, 647, 662

\bibitem[\protect\citeauthoryear{Schmitt \etal}{1996}]{Schmitt1996}
Schmitt, D., Schuessler, M., Ferriz-Mas, A. 1996, 
\aap, 311, L1-L4

\bibitem[\protect\citeauthoryear{Steinhilber \etal}{2012}]{Steinhilber2012}
Steinhilber, F., Abreu., J., Beer, J., \etal 2012,
PNAS, 109, 5967

\bibitem[\protect\citeauthoryear{Steenbeck \etal}{1966}]{Steenbeck1966}
Steenbeck, M., Krause, F., R\"{a}dler, K. -H 1966,
Z. Naturforsch, 21, 1285

\bibitem[\protect\citeauthoryear{Tobias}{1996}]{Tobias1996}
Tobias, S. M. 1996,
\apj, 467, 870

\bibitem[\protect\citeauthoryear{Tobias}{1997}]{Tobias1997}
Tobias, S. M. 1997,
\aap, 322, 1007

\bibitem[\protect\citeauthoryear{Tobias}{2001}]{Tobias2001}
Tobias, S. M., Brummell, N. H., Clune, T. L.; Toomre, J. 2001,
\apj, 549, 1183

\bibitem[\protect\citeauthoryear{Tobias}{2002}]{Tobias2002}
Tobias, S. M. 2002,
Astron. Nachr., 323, 417

\bibitem[\protect\citeauthoryear{Usoskin \etal}{2007}]{Usoskin2007}
Usoskin, I., Solanki, S., Kovaltsov, G. 2007,
\aap, 471, 301

\bibitem[\protect\citeauthoryear{Usoskin}{2008}]{Usoskin2008}
Usoskin, I. 2008,
Living Rev. Solar Phys., 5, 3

\bibitem[\protect\citeauthoryear{Wang \etal}{1991}]{Wang1991}
Wang, Y. -M., Sheeley, N. R., Nash, A. G. 1991,
\apj, 383, 431

\bibitem[\protect\citeauthoryear{Wilmot-Smith \etal}{2006}]{Wilmot-Smith2006}
Wilmot-Smith, A. L., Nandy, D., Hornig, G., Martens, P. C. H. 2006,
\apj, 652, 696

\bibitem[\protect\citeauthoryear{Yeates \etal}{2008}]{Yeates2008}
Yeates, A. R., Nandy, D., Mackay, D. H. 2008,
\apj, 673, 544


\end{thebibliography}
\end{document}